\documentclass[aps,prb,twocolumn,showpacs,final,floatfix]{revtex4-2}
\usepackage{amsmath}
\usepackage{amsfonts}
\usepackage{amssymb}
\usepackage{physics}
\usepackage{color}
\usepackage{hyperref}
\usepackage{inputenc}
\usepackage{enumerate}
\usepackage{graphicx}
\usepackage{float}
\usepackage{amsmath}
\usepackage{amsfonts}
\usepackage{lipsum}
\usepackage{dcolumn}
\usepackage{hyperref}
\usepackage{subfigure}
\usepackage{nicematrix}
\usepackage{epsfig}
\usepackage{epsf} 
\usepackage{epstopdf}
\usepackage[normalem]{ulem}

\usepackage{color}
\usepackage{enumitem}
\usepackage{xr}
\makeatletter

\begin{document}


\title{Quantized two terminal conductance, edge 
states and current patterns in an open geometry 2-dimensional Chern insulator}


\author{Junaid Majeed Bhat$^1$}

\author{R Shankar$^2$}
\author{Abhishek Dhar$^1$}

\affiliation{$^1$International Centre for Theoretical Sciences,  Tata Institute of Fundamental Research, Bengaluru 560 089, India, \\ 
$^2$The Institute of Mathematical Sciences, C I T Campus, Chennai 600 113, India}


\date{\today}

\begin{abstract}
The quantization of the two terminal conductance in 2D topological systems is justified by  the Landauer-Buttiker (LB) theory that  assumes perfect point contacts between  single channel leads and the sample.  We examine this assumption in a microscopic model of a Chern insulator connected to leads, using the nonequilibrium Green’s function formalism. We find that the currents are localized both in the leads and in the insulator and enter and exit the insulator only near the corners. The contact details
do not matter and a  single channel with perfect  contact is emergent, thus
justifying the LB theory. The quantized two-terminal conductance shows interesting finite-size effects and  dependence on system-reservoir coupling.  
\end{abstract}

\maketitle
\section{Introduction:}
\label{sec:intro}
The quantum Hall effect (QHE) was discovered by von Klitzing~\cite{klitzing1980new}
in 1980.  Topological properties were invoked to understand why the Hall
conductance was so exactly quantized in units of $e^2/h$. In a seminal paper,
Thouless and others~(TKNN)~\cite{thouless1982quantized} identified the Chern
invariant, $\nu$, with the Hall conductivity, $\sigma_H=\nu e^2/h$. Hatsugai~\cite{hatsugai1993chern} showed that a non-zero Chern
invariant, $\nu$, of the closed system implies the existence of $\nu$ chiral
edge channels in systems with edges. Chiral edge channels are central to the physics of insulators with non-trivial
topology   in two dimensions. They carry dissipationless current and are, to a certain
degree, robust towards symmetry preserving disorder.  Apart from quantum hall systems, several other insulators with non-trivial
topologies have been discovered, namely those having
non-zero Chern invariants without a magnetic field, the so-called Chern 
insulators (CI)~\cite{PhysRevLett.61.2015,chang2013experimental,jotzu2014experimental}  and time reversal invariant insulators with non-trivial 
$Z_2$ invariants, the so-called topological insulators (TI)~\cite{kane2005,bernevig2006,rahul2009,roy2009,fu2007,moore2007topological,konig2007,hasan2010colloquium}. Topological invariants have also
been defined for closed aperiodic systems such as quasi-crystals and amorphous
systems, characterized by the so-called Bott
index~\cite{huang2018quantum,bandres2016topological,PhysRevLett.118.236402}.
All these systems are characterized by chiral edge channels that provide the
experimental signature of the non-trivial topology. The fact that the edge
channels are dissipationless leads to the possibility of applications in
devices~\cite{tokura2017emergent}. There is growing activity in this field,
so-called topological electronics, which involves engineering the edge
channels.

{
Topological invariants are defined only for closed periodic systems, whereas
experiments are typically done in open systems coupled to leads, where the
current is injected/ejected. In a Hall bar geometry, the probes are away from
the leads. Thus the Hall conductance, $G_H$, and the longitudinal conductance, $G_L$ measured in the Hall bar geometry may not depend on the details of the coupling to the leads. If the TKNN result, derived for a closed CI, were to be applied for the open system, it would predict that $G_H$ is quantized and $G_L=0$.  Decades of experiments on QHE systems have confirmed these predictions.  
For  2-dimensional TI's it has been proposed~\cite{bernevig2006,gusev2019}  that the signal of a  non-trivial $Z_2$ invariant is the quantization of the \emph{two-terminal} conductance, $G$. This has been observed in experiments~\cite{konig2007,roth2009nonlocal,fuchs2023kondo} and simulations~\cite{huang2018quantum,PhysRevLett.118.236402,reinthaler2012interplay}. The proposal was based on theoretical models using the Landauer-Buttiker (LB) formalism, with additional assumptions on  perfect transmission  via point contacts from  single channel leads to the edge channels~\cite{bernevig2006,dolcetto2016,gusev2019}(the arguments of \cite{dolcetto2016} are reproduced in App.~\ref{app:equi}).  It is not theoretically obvious that the assumption of  single channel leads and perfect transmission via point contacts is physically valid.  In fact it has so far not been shown that this assumption follows from a microscopic approach such as the non-equilibrium Green’s function (NEGF) formalism, where the leads are coupled to the system across its width and the transmission is explicitly related to details of both the reservoir model and its coupling with the system. This is the “theory gap” that is addressed in this work.  Note that the equivalence between multi-channel LB,  NEGF and the Green-Kubo formalism has been established~\cite{fisher1981relation,dmitry2016theory}. However, one cannot use the TKNN method to prove quantization  since we now have an aperiodic inhomogeneous system,  while the TKNN proof holds only for periodic homogeneous systems.

Since the topological properties of a 2-dimensional TI can be modeled as those of two decoupled CI's with opposite Chern numbers, we consider the strip geometry of a CI connected to metallic leads at two opposite edges, with a voltage difference V applied across the leads. Here one can measure not only $G_H$ but also the two-terminal longitudinal conductance $G$, which is the main focus of this paper.  Within the LB formalism it is easy to show that  $G$ is the same as $G_H$~\cite{gusev2019}, see Appendix ~\ref{app:equi}.}

 In this work, we attempt to arrive at a better understanding of the
two-terminal longitudinal conductance, $G$, in the open system by use of the
NEGF formalism.  For our studies, we consider the spinless BHZ~(SBHZ)
model~\cite{shankar2018}, a Chern insulator, placed in contact with two 
metallic leads.  Apart from measuring  the conductance obtained from NEGF, we
use this formalism to also extract information on the scattering states formed
by the edge modes \emph{in the presence of the leads}. The strip geometry makes
this a highly non-trivial problem. In particular, for the scattering states, we
obtain the current and charge density profiles inside  the insulating region as
well as in the metallic leads.

 We summarize our main findings: 
\begin{enumerate}
	\item  From our numerics, based on NEGF,  we verify that the two-terminal longitudinal conductance, $G$, is quantized when the Fermi level is in the band gap of the insulator. For sufficiently large  system sizes, this is independent of the strength of the coupling at the contacts between the system and leads. We also look at finite size effects of the quantized two-terminal conductance and find that the growth of the conductance, to the quantized value, shows damped oscillations as a function of both the system size and the system-reservoir coupling.  The oscillation period shows a simple scaling with the system size and the coupling strength. 
		\item    We generalize the NEGF approach to compute currents both in the CI and in the metallic leads. We find that in the insulating region,  the current density due to the edge modes  is, as expected, localized along the boundaries of the sample. Remarkably, for the case when the reservoirs are in the vicinity of vanishing Fermi levels, we find that  even inside the leads, the current density is highly localized and moves along zig-zag lines at  $45^\circ$ to the longitudinal direction. The current
	only enters and leaves the insulator near the  diagonally opposite  corners,  despite
	the fact that the reservoirs are coupled to the insulator throughout its
	width.  This justifies the emergence of the perfect point contact, which is the main assumption of the LB formalism.  As the Fermi level is shifted from the middle of the 
	gap, the current density is less sharply localized but the injection(ejection)
	to(from) the CI is still near the corners. To understand the corner injection/ejection, we present some analytic results for the zero energy wavefunctions of a metal-CI junction with periodic boundary conditions along the junction. 
\end{enumerate}

 The rest of the paper is structured as follows: In Sec.~\ref{sec:model_negf}, we present the details of the Hamiltonian of the CI and the open system geometry considered in this paper. In the next section, Sec.~\ref{sec:neg_form}, we set up the  NEGF formalism for computation of current and charge density in the CI and the leads. Sec.~\ref{sec:numer} presents numerical results on the conductance of the CI and current density patterns inside the leads and the CI. We also look at the effects of deforming the shape of the Fermi-surface of the reservoir and contact disorder on the observed current density patterns. In the next section, Sec.~\ref{sec:mci_jun}, we provide an argument for the phenomenon of corner injection by considering a Metal-CI junction and showing that no current is injected into the CI from the reservoir at the junction. We conclude in Sec.~\ref{sec:concl}.
\section{The Model}
\label{sec:model_negf}
 The SBHZ model is a simple 2D topological insulator given by a nearest neighbour tight-binding Hamiltonian on an $N_x\cross N_y$ rectangular sample.  At any lattice point $(x,y)$, there are two fermionic degrees of freedom described by the annihilation operators $\psi_1(x,y)$  and $\psi_2(x,y)$, respectively. The Hamiltonian of the insulator is given by, 
 \begin{align}
\label{HW-CI}
&\mathcal{H}_W=\sum_{x,x'=1}^{N_x}\sum_{y,y'=1}^{N_y}  \Psi^\dagger(x,y)H_W[x,y;x',y']\Psi(x',y'),\\
&=\sum_{x=1}^{N_x}\sum_{y=1}^{N_y}\mu_w\Psi^\dagger(x,y)\sigma_z\Psi(x,y)\notag\\&+\sum_{x=1}^{N_x-1}\sum_{y=1}^{N_y}\left[\Psi^\dagger(x,y)\frac{1}{2}(\sigma_z+i\sigma_x)\Psi(x+1,y)+\text{h.c.}\right]\notag\\&+\sum_{x=1}^{N_x}\sum_{y=1}^{N_y-1}\left[\Psi^\dagger(x,y)\frac{1}{2}(\sigma_z+i\sigma_y)\Psi(x,y+1)+\text{h.c.}\right], \nonumber
\end{align}
 where we have defined the two-vectors $\Psi(x,y)=\begin{pmatrix}\psi_1(x,y)&&\psi_2(x,y)\end{pmatrix}^T$. The topologically non-trivial phases of Chern number 1 and -1 lie in the parameter regimes of $-2<\mu_w<0$  and $0<\mu_w<2$, respectively. In these parameter regimes, the rectangular sample with edges  supports dissipationless edge modes  with energies that lie within the gap of the insulator.  

   \begin{figure*}
	\subfigure[]{\includegraphics[width=0.45\textwidth]{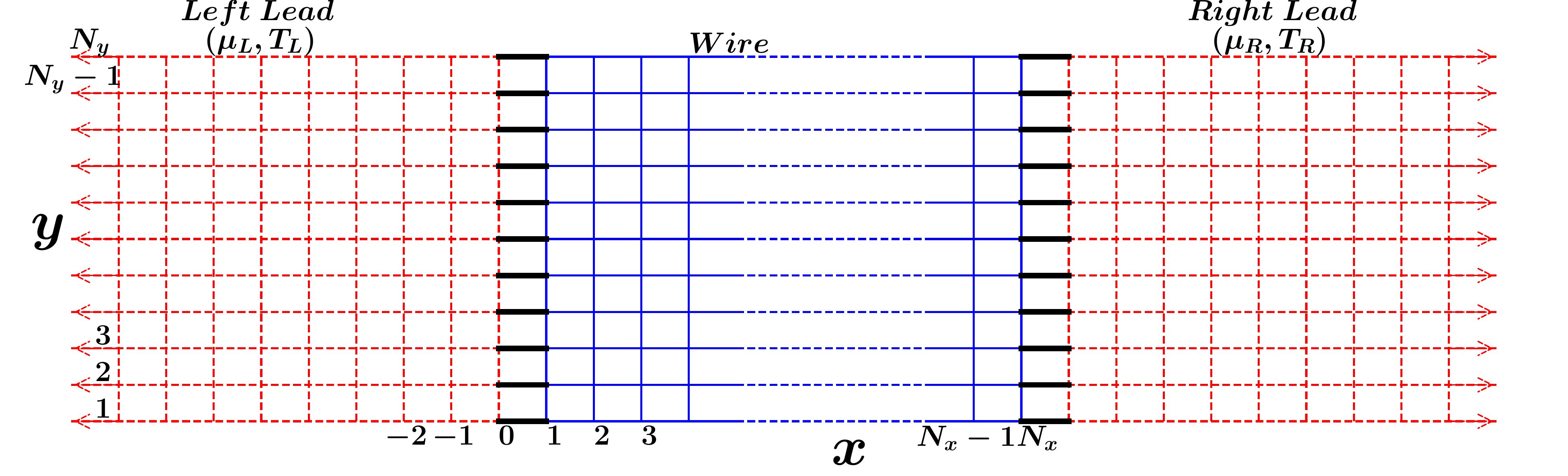}}
	\subfigure[]{\includegraphics[width=0.45\textwidth]{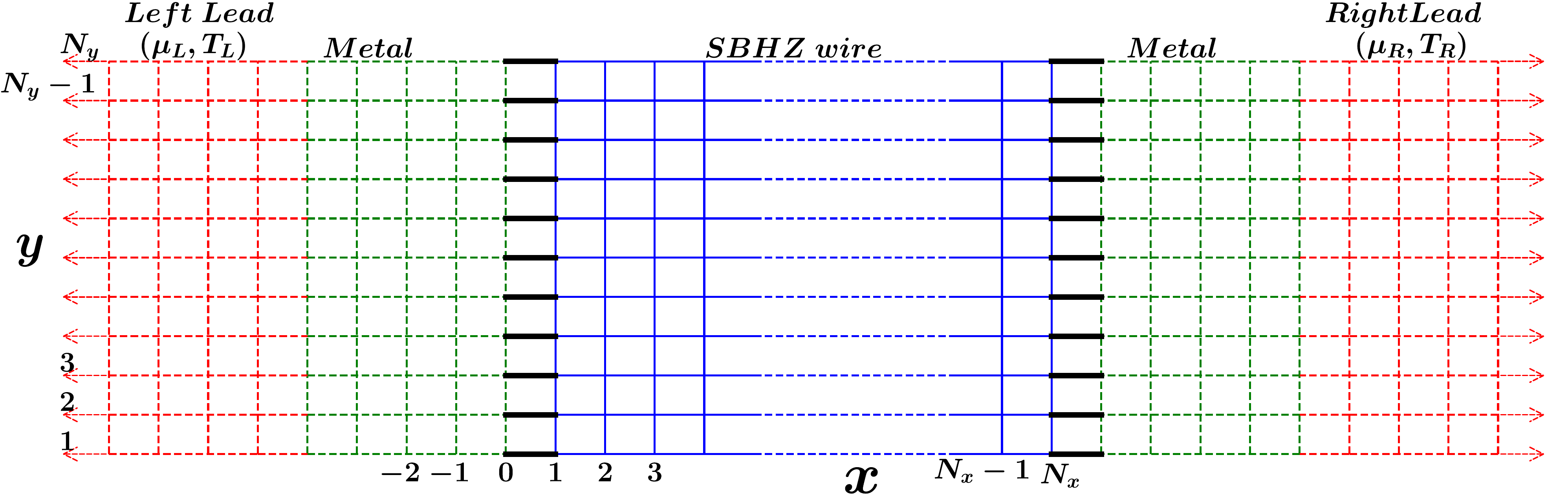}}
	\caption{Schematic of two geometries used for the calculation of the current density in the SBHZ wire and the reservoirs. The red regions are the reservoirs. The green region is the normal metallic region with a Hamiltonian identical to the reservoirs. The blue region is the SBHZ wire. The microscopic Hamiltonian for the two geometries is exactly the same  and the two only differ   in the initial condition, namely the red regions where the temperatures and chemical potentials are specified differ in the two geometries.}
	\label{fig:schsup}
\end{figure*} 
 To study  transport across the CI via the edge modes, we consider   the system in contact with two external reservoirs at the two opposite edges at $x=1$ and $x=N_x$. We show in Fig~(\ref{fig:schsup}a) a schematic of the setup that we consider.  Each reservoir is taken to consist of two decoupled layers of  metallic leads that are semi-infinite in the $x$-direction and of width $N_y$ in the $y$-direction and are modeled as 2D nearest neighbour tight-binding Hamiltonians with hopping $\eta_b$. Thus we take two fermionic degrees of freedom on each site in the two reservoirs as well. We label the corresponding annihilation operators as $\psi_{1,L/R}(x_{L/R},y)$ and $\psi_{2,L/R}(x_{L/R},y)$, where $L$ and $R$ label the left and the right lead, respectively. We then define the two-vectors $\Psi_{L/R}(x,y)=\begin{pmatrix}\psi_{1,L/R}(x,y)&&\psi_{2,L/R}(x,y)\end{pmatrix}^T$. The integers $x_L$ and $x_R$ label the $x$ coordinates of the sites on the left and the right reservoirs, respectively. These take integer values  that run from $-\infty$ to $0$ and $N_x+1$ to $\infty$, respectively. The $y$ coordinate here takes the same integer values as in the wire~(SBHZ insulator) i.e. from $1$ to $N_y$. Free boundary conditions are imposed at the edges of the reservoir and the system at $y=1$ and $y=N_y$ respectively. The full Hamiltonian of the system and baths is therefore given by:
\begin{equation}
\mathcal{H}=\mathcal{H}_L+\mathcal{H}_{WL}+\mathcal{H}_W+\mathcal{H}_{WR}+\mathcal{H}_R.
\end{equation}
where $\mathcal{H}_W$ is the Hamiltonian of the SBHZ CI  in Eq.~(\ref{HW-CI}) while  the other terms correspond to the reservoir Hamiltonians, 
\begin{widetext}
\begin{align}
	\mathcal{H}_{L}&=\sum_{x_L,x_L'=-\infty}^{0}\sum_{y=1}^{N_y} \Psi^{\dagger}_L(x_L,y)H_L[x_L,y;x'_L,y']\Psi_L(x'_L,y') \notag\\
	&=\eta_b\sum_{x_{L}=-\infty}^{-1}\sum_{y=1}^{N_y}\left[ \Psi_{L}^\dagger(x_L,y)\Psi_L(x_{L}+1,y)+\text{h.c.}\right]+\eta_b\sum_{x_{L}=-\infty}^{0}\sum_{y=1}^{N_y-1}\left[\Psi_{L}^\dagger(x_L,y)\Psi_L(x_{L},y+1)+\text{h.c.}\right],\\
	\mathcal{H}_{R}&=\sum_{x_R,x_R'=N_x+1}^{\infty}\sum_{y=1}^{N_y} \Psi^{\dagger}_R(x_R,y)H_R[x_R,y;x_R',y']\Psi_R(x'_R,y') \notag\\
	&=\eta_b\sum_{x_{R}=N_x+1}^{\infty}\sum_{y=1}^{N_y}\left[ \Psi_{R}^\dagger(x_R,y)\Psi_R(x_{R}+1,y)+\text{h.c.}\right]	+\eta_b\sum_{x_{R}=N_x+1}^{\infty}\sum_{y=1}^{N_y-1}\left[ \Psi_{R}^\dagger(x_R,y)\Psi_R(x_{R},y+1)+\text{h.c.}\right], 
\end{align}
and the system-reservoir coupling Hamiltonians, 
\begin{align}
	\mathcal{H}_{WL}&=\sum_{x_L=-\infty}^0\sum_{x=1}^{N_x}\sum_{y,y'=1}^{N_y}\left[\Psi_{L}^\dagger(x_L,y)V_L[x_L,y;x,y']\Psi[x,y']+\text{h.c.}\right]=\eta_c\sum_{y=1}^{N_y} \left[\Psi^\dagger_L(0,y)\Psi(1,y) +\text{h.c.}\right],\\
	\mathcal{H}_{WR}&=\sum_{x_R=N_x+1}^\infty\sum_{x=1}^{N_x}\sum_{y,y'=1}^{N_y}\left[\Psi_{R}^\dagger(x_R,y)V_R[x_R,y;x,y']\Psi[x,y']+\text{h.c.}\right]=\eta_c\sum_{y=1}^{N_y} \left[ \Psi^\dagger_R(N_x+1,y)\Psi(N_x,y') +\text{h.c.}\right],
\end{align}
\end{widetext}
where $\eta_c$ is the coupling  between the CI and the reservoirs.

\section{NEGF Formalism}
\label{sec:neg_form}
Using NEGF one can obtain the non-equilibrium steady state of this system starting from a state where the left and the right reservoirs are described initially by grand canonical ensembles with chemical potentials and temperatures $(\mu_L,T_L)$ and $(\mu_R,T_R)$ while the wire  is in an arbitrary state.  The non-equilibrium steady state solution for the wire operators in the limit $t\rightarrow \infty$ can be derived in terms of the effective Green's function of the wire given by $\mathcal{G}(\omega)=(\omega - H_W-\Sigma_L(\omega)-\Sigma_R(\omega))^{-1}$, where
$\Sigma_{L/R}(\omega)~$ are the self energies from the reservoirs. From this solution, particle and heat currents and, in fact, all two-point correlators can be derived~\cite{dhar2006,PhysRevB.102.224512}.  Here we are specifically interested in the conductance, and the profiles of  charge density and the current density in the wire (including in the reservoirs) for the zero-temperature case, in the linear response regime ($\mu_L=\mu+\Delta\mu_L$ and $\mu_R=\mu+\Delta\mu_R$, with  $\Delta \mu_L,\Delta \mu_R << \mu$).

The total current passing through the CI, in units of $e^2/h$, is given by  $I= G(\mu)(\Delta\mu_L-\Delta \mu_R)$, where
$G(\mu)=4\pi^2\Tr[\mathcal{G}(\mu)\Gamma_R(\mu)\mathcal{G}^{\dagger}(\mu)\Gamma_L(\mu)]$ is the two-terminal conductance of the CI  and $\Gamma_{L/R}(\mu)=[\Sigma_{L/R}(\mu)-\Sigma_{L/R}^\dagger(\mu)]/(2\pi i)$. 

 The expressions for the  current density  can be computed from the correlation matrix, $C$ with components given by
$ C_{\mathbf{x} ,a;\mathbf x' ,b}=\expval{\psi_a^\dagger(\mathbf x)\psi_b(\mathbf x')}\label{correlators}$ with $a,b= 1,2$, $\mathbf{x}=(x,y)$ and   the expectation value is taken in the non-equilibrium steady state. Using NEGF, $C$ can be expressed in terms of the effective Green's function $\mathcal{G}(\omega)$ and the Fermi functions of the reservoirs as~\cite{dhar2006,PhysRevB.102.224512},
\begin{align}
C=&\int_{-\infty}^{\infty} d\omega~[ C^L(\omega)f(\omega,\mu_L,T_L)+C^R(\omega)f(\omega,\mu_R,T_R)]\label{denexp},
\end{align}
where $f(\omega,\mu,T)$ is the Fermi Dirac distribution function, $C^{L/R}(\omega)=2\pi\mathcal{G}(\omega)\Gamma_{L/R}(\omega)\mathcal{G}^\dagger(\omega)$ and $\Gamma_{L/R}=(\Sigma_{L/R}^\dagger-\Sigma_{L/R})/(2\pi i)$. The local charge density is given  terms of $C$ as follows:
\begin{align}
\rho(\mathbf{x})= \sum_{a=1}^2 C_{\mathbf{x},a;\mathbf{x},a}\label{eq:chargeden},
\end{align}
while the net particle current  on the bond between the lattice points $\mathbf x$ and $\mathbf x'$ is given by  
\begin{align}
J_{\mathbf x,\mathbf x'}=-2\sum_{a,b=1}^2\Im\bigg[H_W[\mathbf x,a;\mathbf x',b] C_{\mathbf x,a;\mathbf x',b}\bigg].\label{currden} 	
\end{align}
 Let us again consider $T=0$ and  the linear response regime. We expect non-vanishing edge currents  even for the case of zero applied voltage, whenever the CI is in a topological phase. It is useful to distinguish these from the nonequilibrium transport current in the presence of a voltage bias. Hence it is useful to write the  current on any bond between sites ${\bf x}=(x,y)$ and ${\bf x'}=(x',y')$ in the following form: 
\begin{align}
J_{\bf x,\bf x'}&=J^{\rm eq}_{\bf x,\bf x'} +\bar{J}^{{\rm L}}_{\bf x,\bf x'}+\bar{J}^{{\rm R}}_{\bf x,\bf x'}, {\rm where} \\
J^{\rm eq}_{\bf x,\bf x'}(\mu)&= \int_{-\infty}^\mu d \omega [F^{\rm L}_{\bf x,\bf x'}(\omega)+F^{\rm R}_{\bf x,\bf x'}(\omega)], \\
\bar{J}^{{\rm L}}_{\bf x,\bf x'}&= F^{\rm L}_{\bf x,\bf x'}(\mu) \Delta \mu_L ,~~ \bar{J}^{{\rm R}}_{\bf x,\bf x'}=F^{\rm R}_{\bf x,\bf x'}(\mu)  \Delta \mu_R,
\end{align}
where the functions $F^{\rm L,R}_{\bf x,\bf x'}$ are defined as,
\begin{align}
	F^{\text{L}/\text{R}}_{\mathbf x,\mathbf x'}(\omega)=-2\sum_{a,b=1}^2\Im\bigg(H_W[\mathbf x,a;\mathbf x',b] C^{L/R}_{\mathbf x,a;\mathbf x',b}(\omega)\bigg).
\end{align}
We call $\bar{J}_{\bf x,\bf x'}=\bar{J}^{{\rm L}}_{\bf x,\bf x'}+\bar{J}^{{\rm R}}_{\bf x,\bf x'}$ as the excess current density  and $I=\sum_{y=1}^{N_y} \bar{J}_{x,y;x+1,y}$ would give the nonequilibrium transport current across the CI.  For points in the metal the equilibrium current, $J^{ eq}_{\bf x,x'}(\mu)=0$ for any $\mu$, which implies that $ F^{ \text{L}}_{\bf x,\bf x'}(\mu)=-F^{ \text{R}}_{\bf x,\bf x'}(\mu)$. However for points inside the CI, the topological phases support edge currents even in equilibrium, meaning that   $J^{\rm eq}_{\bf x,\bf x'}(\mu)$ is non-vanishing on the edges, consequently   $ F^{L}_{\bf x,\bf x'}(\mu)\neq -F^{ R}_{\bf x,\bf x'}(\mu)$. This leads to the interesting observation that the excess current density in the CI does not simply change sign when we interchange $\Delta \mu_L$ and  $\Delta \mu_R$ (though the total current  $I=G (\Delta \mu_L-\Delta \mu_R)$ has this property), and therefore is sensitive to the choices of $\Delta\mu_L$ and $\Delta\mu_R$.   Note  that the total  equilibrium current across any transverse cross-section vanishes, as expected in equilibrium.  We emphasize that the excess current, $\bar{J}$, is a relatively small correction over  ${J}^{ eq}$, and  is the transport current arising due to the chemical potential difference between the reservoirs. Only the total current density is expected to respect the chirality of the Chern insulator, while the excess current density can flow opposite to the chirality of the insulator since it is proportional to $\Delta\mu_{L/R}$.

The charge density can also be split as the sum of equilibrium part and an excess part as, $\rho(\mathbf{x})=\rho^{eq}(\mathbf x)+\bar \rho(\mathbf{x})$ where,
\begin{align}
	&\rho^{eq}=\sum_{a=1}^2\int_{-\infty}^\mu d\mu\left[ C^L_{\mathbf{x},a,\mathbf{x},a}(\mu) + C^R_{\mathbf x,a;\mathbf x',a}(\mu)\right]\\
	&\bar \rho(x,y)=  \sum_{a=1}^2C^L_{\mathbf{x},a,\mathbf{x},a}(\mu)\Delta\mu_L + C^R_{\mathbf x,a;\mathbf x',a}(\mu)\Delta\mu_R.\label{charden}
\end{align}
\begin{figure*}
	\subfigure{\includegraphics[width=0.3\textwidth,height=0.19\textwidth]{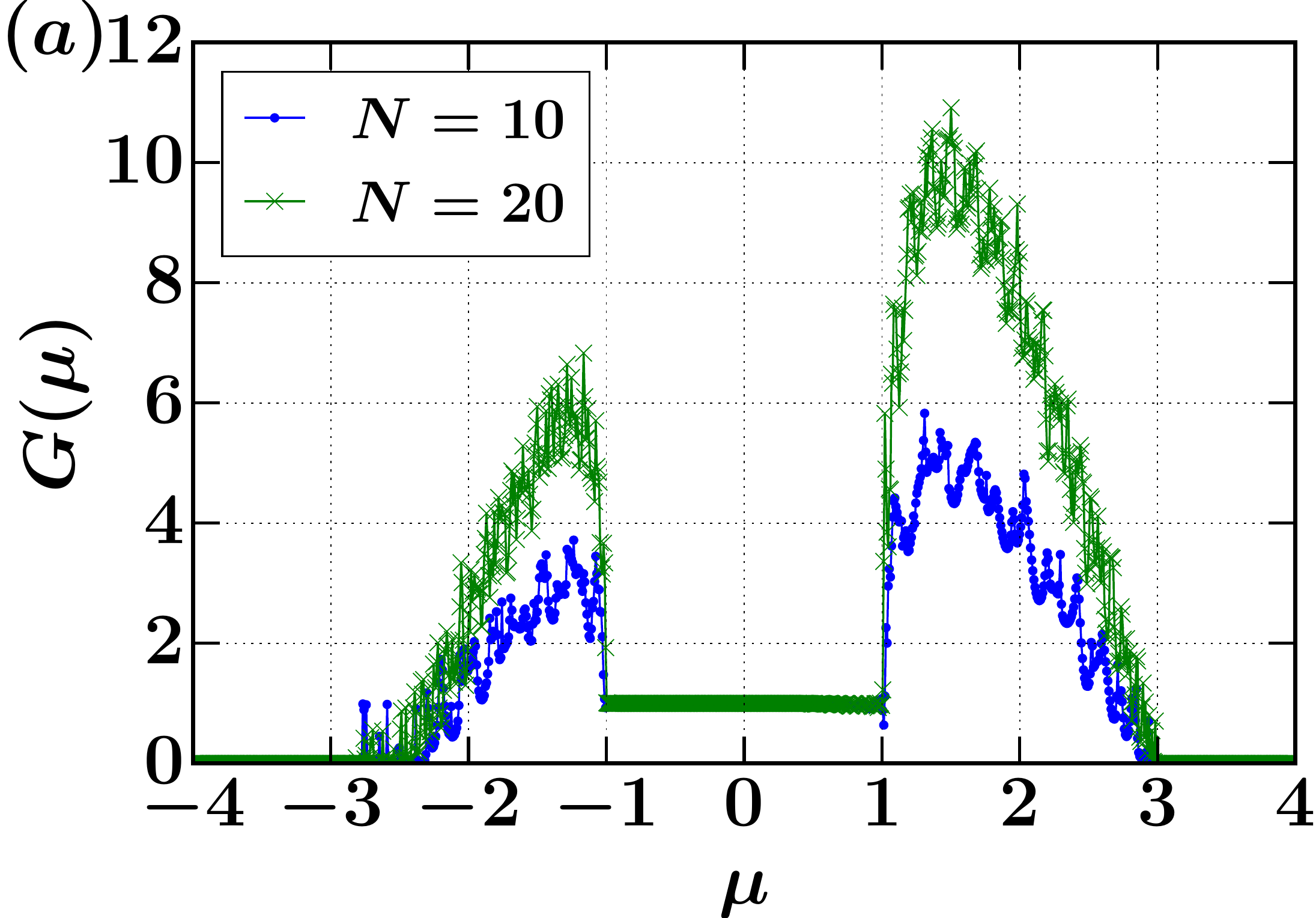}}
	\subfigure{\includegraphics[width=0.3\textwidth,height=0.19\textwidth]{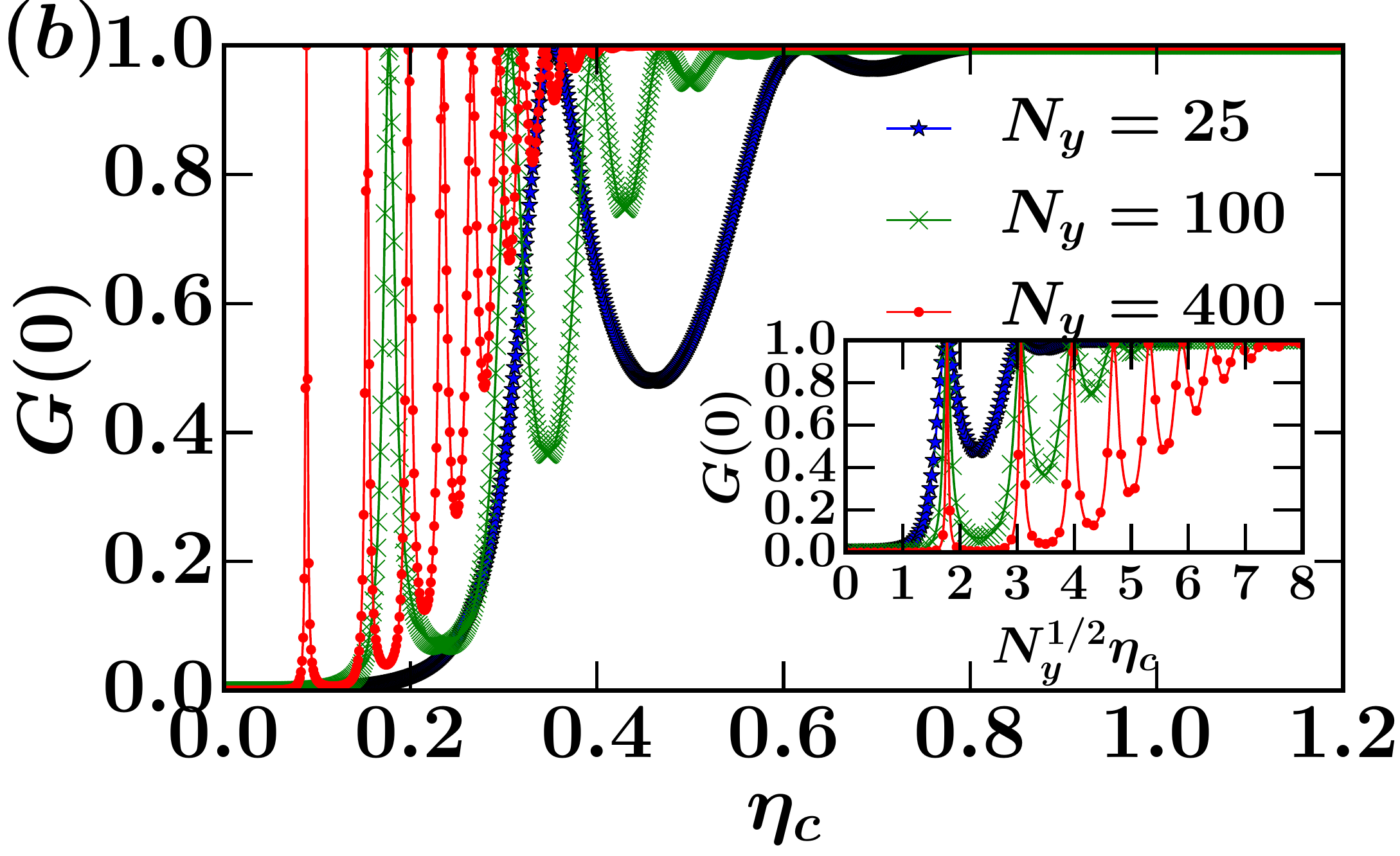}}
	\subfigure{\includegraphics[width=0.3\textwidth,height=0.19\textwidth]{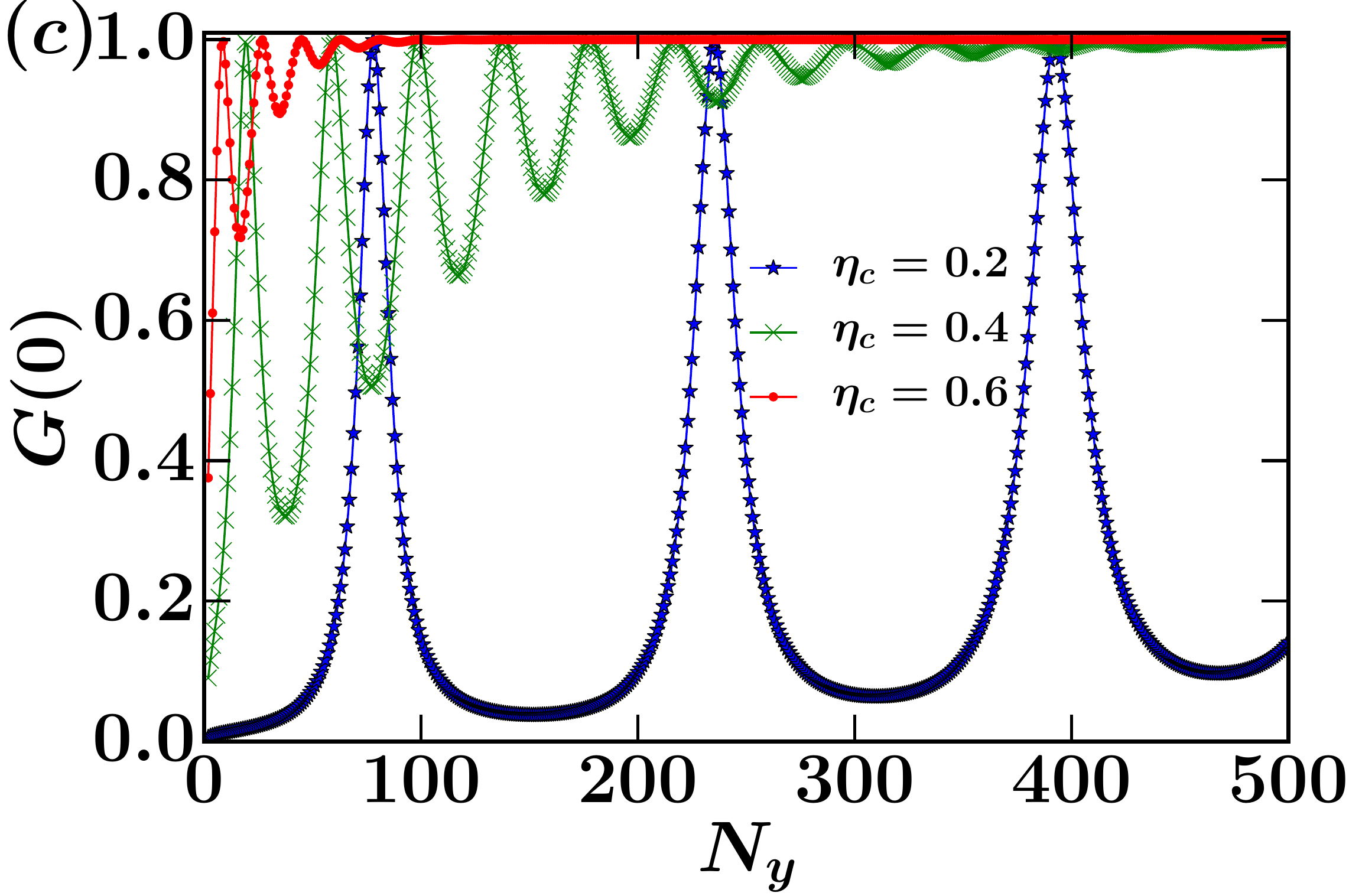}}
	\caption{In (a) we show the variation of the conductance with the Fermi level, $\mu$.  (b)  and (c)  show the variation of the conductance at the Fermi level, $\mu=0$,  with  $\eta_c$ and $N_y$ at $N_x=100$, respectively. The inset in (b) shows that, on scaling $\eta_c$ with $N_y^{1/2}$, the peaks in the oscillations coincide and therefore the period of the oscillations scales as $N_y^{1/2}\eta_c$. Parameter values:  (a) $N_x=N_y=N$, $\eta_b=2$, $\mu_w=1$, $\eta_c=1.5$ and for (b) and (c) $N_x=100$ $\eta_b=2$ and $\mu_w=1$.}	
	\label{fig:cond}
\end{figure*}
\begin{figure*}
	\centering
	\subfigure{\includegraphics[width=\textwidth]{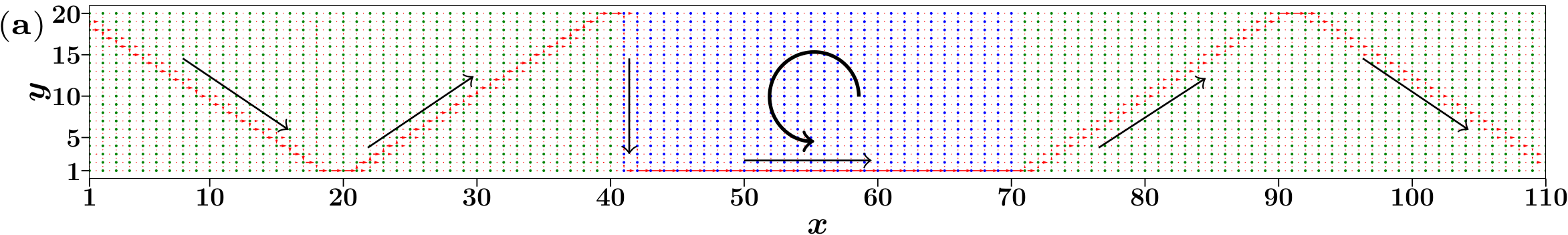}}
	\subfigure{\includegraphics[width=\textwidth]{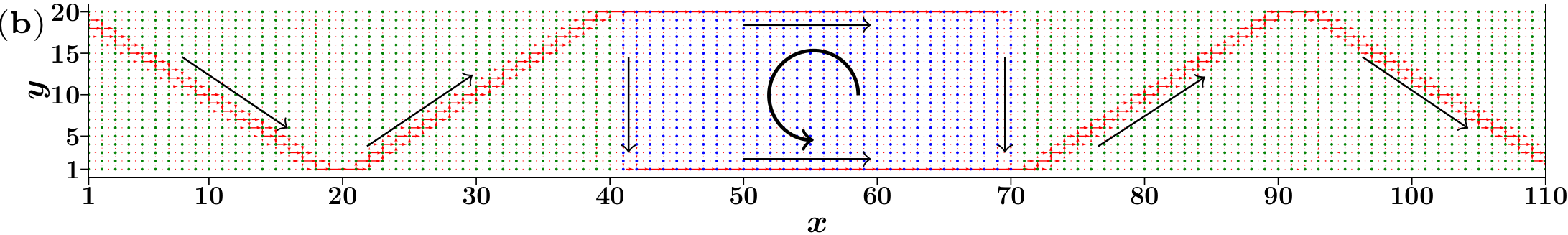}}
	\caption{Excess current density in the metallic leads and the SBHZ wire at zero Fermi level with (a) $\Delta\mu_R=0~,\Delta\mu_L>0$ and (b) $\Delta\mu_L=-\Delta\mu_R=\Delta\mu$. The current in the reservoirs is localized along the lines at $\pm 45$ degrees to the horizontal directions and enters/leaves the CI near the corners.  The black straight arrows indicate the direction of the flow of the current and the curved arrow indicates the chirality of the CI.  The pattern of the current inside the CI in  (a)  is different from (b), illustrating that the excess current density patterns are sensitive to the choices of $\Delta\mu_L$ and $\Delta\mu_R$. Parameter values: $N_x\times N_y=30\times20$ and $\eta_b=\eta_c=\mu_w=1$.}
	\label{fig:cden1}
\end{figure*}

While Eq.~(\ref{eq:chargeden}) and Eq.~(\ref{currden})  gives us the charge and the current density  inside the system,  these can also be used to compute the current density in the leads with a slight modification of $H_W$. The main point to note is that in the NESS of the full system and reservoirs, the state inside the reservoirs has  also evolved and in fact contains information about the scattering states (plane waves incident from either reservoir onto the scattering region formed by the insulator).    We thus  consider a  setup (see Fig~(\ref{fig:schsup}b))  where now the wire consists of  a CI (blue region) that is sandwiched between finite metallic segments (green region), which are in turn connected to the metallic reservoirs (red region).  Compared to the original setup in Fig~(\ref{fig:schsup}a), the new setup now includes parts of the lead Hamiltonians (green region) inside the system Hamiltonian, $H_W$. Therefore,  we can now use the NEGF formalism to compute current everywhere inside the system, including the green metallic sites. The current density in any finite segment of the two setups in  Fig~(\ref{fig:schsup}) is  identical  because of  uniqueness of the NESS.   We now present numerical results for the conductance $G(\mu)$ and the excess  current density profiles inside the system as well as in the leads.

\begin{figure*}	
	\subfigure{\includegraphics[width=0.9\textwidth]{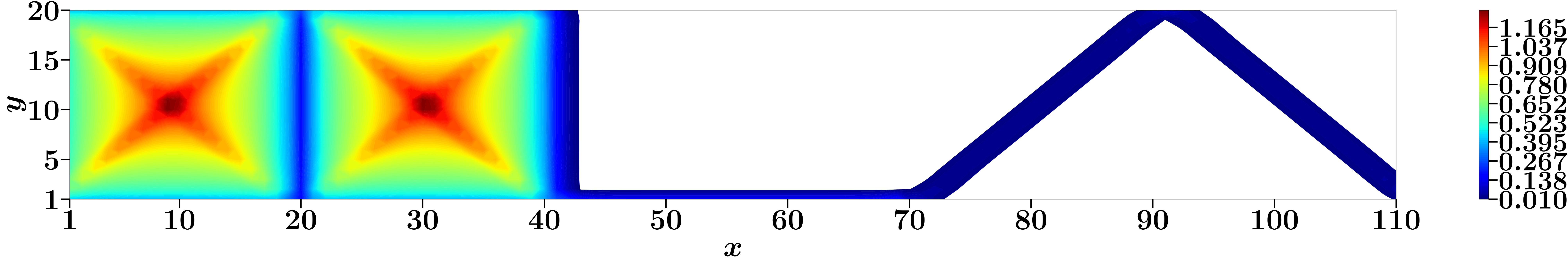}}
	\caption{Excess charge in the leads and the SBHZ wire at zero Fermi level.  Deep inside the insulator (white region), the density has a value  less than $10^{-8}$. Parameter values--$N_x\times N_y=30\times20$, $\Delta\mu_R=0$, $\eta_b=\eta_c=1$ and $\mu_w=1$.}
	\label{fig:cden_sup}	
\end{figure*}

\section{Numerical Results}
\label{sec:numer}
\subsection{Conductance, Charge density and Current density}

In Fig.~(\ref{fig:cond}), we plot the conductance as a function of the Fermi level, $\mu$, for two system sizes. It is seen that  when the Fermi level lies in the gap, the value of the conductance  (in units of $e^2/h$) is quantized to the value $G=1$.  In Fig.~(\ref{fig:cond}b,\ref{fig:cond}c) we show the variation of the conductance, at Fermi level $\mu=0$,   with the strength of the coupling with the reservoirs, $\eta_c$, and the width of the insulator, $N_y$, respectively.  
 From these plots, we see that the conductance strength approaches the quantized value and  eventually  becomes independent of $\eta_c$ and $N_y$. The approach to the quantized value is oscillatory, with exact quantization being achieved  at some specific values of $\eta_c$ and $N_y$. The inset of Fig.~(\ref{fig:cond}b) shows, remarkably, that the oscillation period scales as $N_y^{1/2}\eta_c$.

Next, in Fig.~(\ref{fig:cden1}) and Fig.~(\ref{fig:cden_sup}), we show the excess current density and charge density for the case with  Fermi level, $\mu=0$,  and $\Delta\mu_L>0$ and $\Delta\mu_R=0$.  We see that the current flows along the edges of the insulator, and the charge density is also localized along the edges~[See Fig.~\ref{fig:cden_sup}], as expected. Surprisingly,  the excess current density is  sharply localized  even in the normal metallic regions.  We see that the current primarily flows along the lines at $45$ degrees to the horizontal direction and gets multiply reflected until it reaches the top corner of the SBHZ wire. At this corner, it gets injected into the insulator, flows along the  edges and then leaves it at the diagonally opposite corner into the normal metallic region on the other end. In Fig.~\ref{fig:cden_sup} we see that the excess charge from the left lead gets distributed into the right lead along the lines of localization of the excess current.

 Fig.~(\ref{fig:cden1}b)  shows the current density pattern for $\Delta\mu_L=-\Delta\mu_R=\Delta\mu$. The current gets injected and ejected from the  diagonally opposite corners as in Fig.~(\ref{fig:cden1}a). However, it now flows along all edges of the CI  which is different from Fig.~(\ref{fig:cden1}a) where it flows only along two of the adjoining edges of the CI. Therefore, the current density pattern in the CI shows nontrivial differences for different choices of $\Delta\mu_L$ and $\Delta\mu_R$. Note that the current density along the opposite edges have opposite chirality, illustrating that the excess current can flow opposite to the chirality of the CI. 
\begin{figure}
	\subfigure{\includegraphics[width=0.49\textwidth]{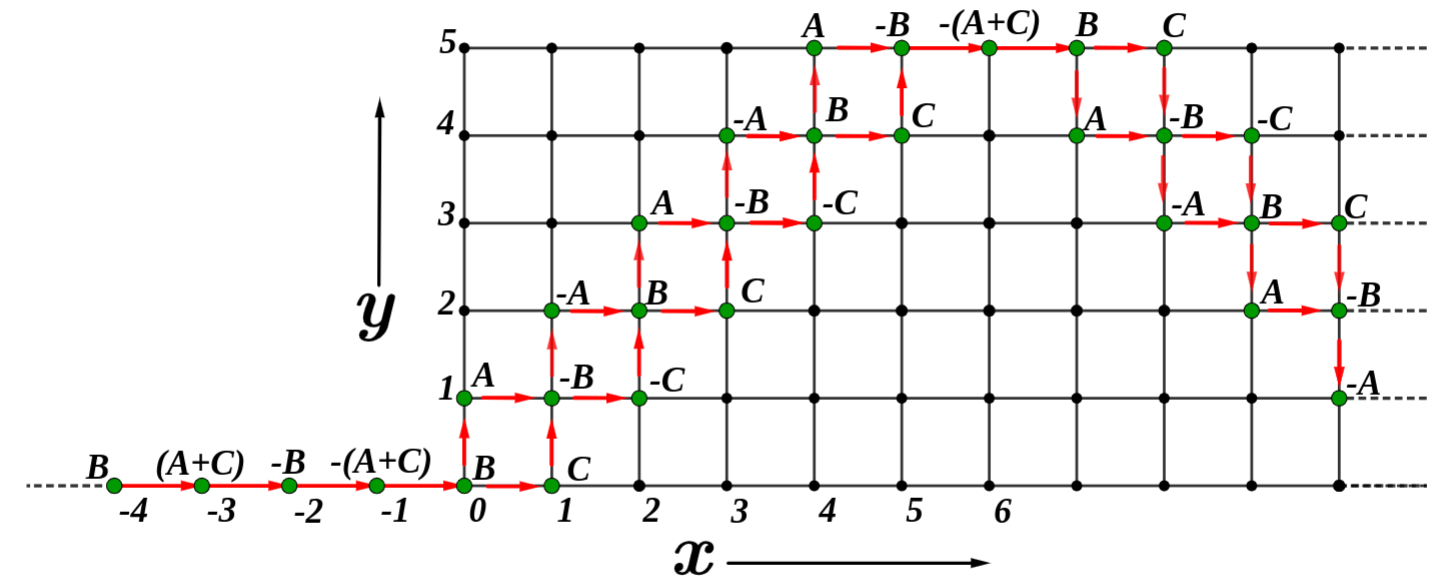}}
	\caption{A scattering state at zero energy in a 2D metallic strip with injection at the left bottom corner. The wavefunction is non-zero only on the green-marked sites, with the amplitudes indicated by $A,B,C$ and their combinations. The red arrows indicate the currents.}
	\label{fig:waf}
\end{figure}

The features of the current density (corner injection and localization in the leads) are sharpest when the  Fermi surface  of the metal is a square in momentum space and smoothly fade away as the Fermi surface is deformed.  In  the next subsection, we illustrate this by presenting numerical results on the effects of deforming the Fermi surface  on the current density.  
The fact that the observed  localization is sharp at $\mu=0$  seems to suggest that it is very special to the scattering states formed  by the  edge modes of the insulator and the reservoir modes   at zero energy. At zero energy, the Fermi surface in the metal is a square in  momentum space. The localization arises from a particular superposition of the  modes on this surface.  To illustrate this, consider the setup in Fig.~\ref{fig:waf} where a current is injected into a semi-infinite metallic strip  of width $W$ at the left-bottom corner through a point contact with a 1D reservoir. The motivation for considering such a setup comes  from the numerical observation (in Fig.~\eqref{fig:cden1}) that the current is injected near the corner into the metallic region. For simplicity, let the Hamiltonian  of the 1D reservoir and the metallic strip   be  of the nearest neighbour tight-binding form, with all the hopping parameters, including that of the  point contact at the corner, set to  $1$. At any lattice site $(x,y)$, the  wave function components of the nearest neighbour points will  sum to zero. In  Fig~\ref{fig:waf}, a solution  is presented for one such  scattering state  with the wave function being non-zero only at certain points (large green circles).  As $W\rightarrow\infty$, the solution can be written as a simple superposition  of states on the square Fermi surface. This state displays a   current localization similar to what  we see numerically in Fig.~(\ref{fig:cden1}b).   
		\begin{figure*}
		\subfigure{\includegraphics[width=0.9\textwidth]{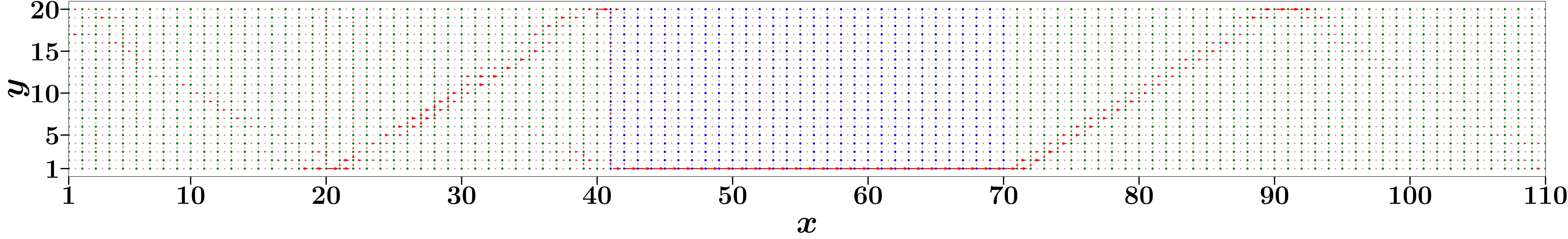}}
		\caption{Excess  current density, ~$\bar{J}/\Delta\mu_L$, in the reservoirs and the SBHZ wire with $\mu=0.2$.   The current in the reservoirs is localized along the lines at $45$ degrees to the horizontal directions. Parameter values-$N_x\times N_y=30\times20$, $\Delta\mu_R=0$, $\eta_b=\eta_c=1$ and $\mu_w=1$.}
		\label{figsupp:cden1}
	\end{figure*}
	\begin{figure*}
		\centering
		\subfigure[~$\bar{J}/\Delta\mu_L$ with $\eta_{by}=1.1$]{\includegraphics[width=0.881\textwidth]{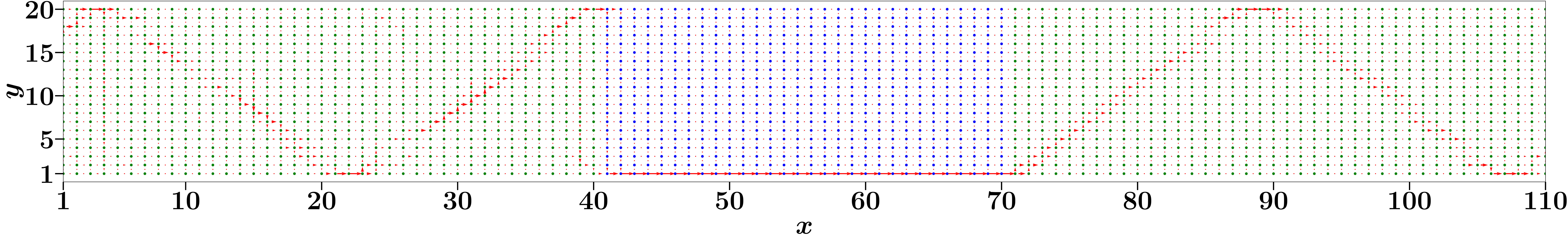}}
		\subfigure[~$\bar{J}/\Delta\mu_L$  with $\eta_{by}=1.2$]{\includegraphics[width=0.881\textwidth]{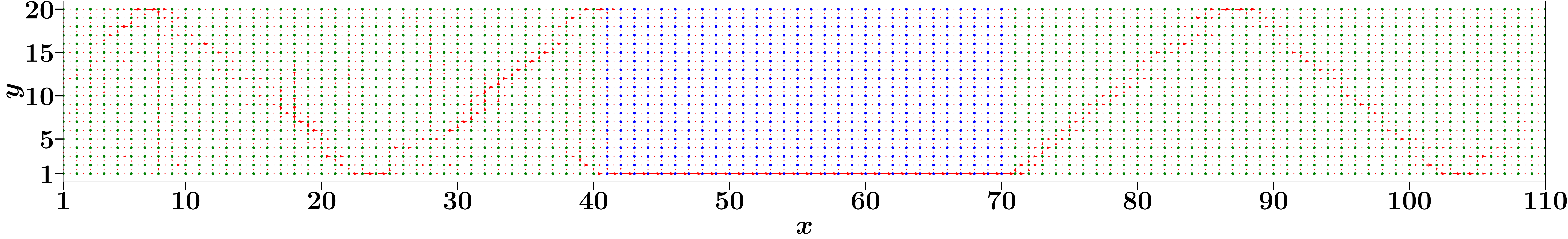}}
		\subfigure[~$\bar{J}/\Delta\mu_L$  with $\eta_{by}=1.3$]{\includegraphics[width=0.881\textwidth]{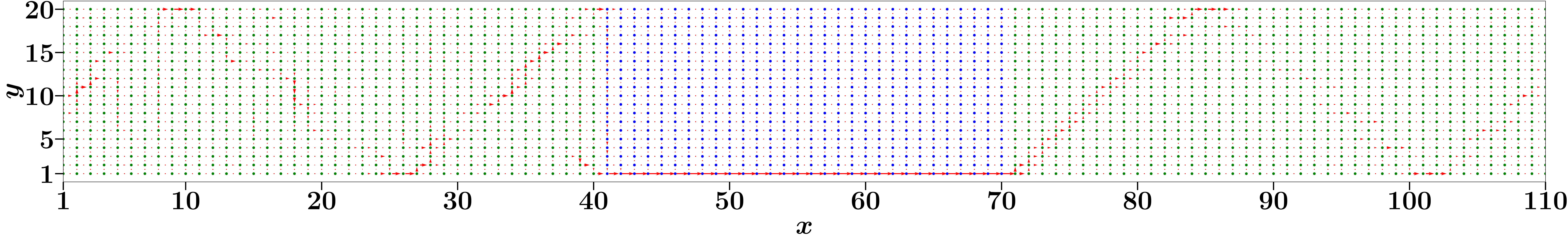}}
		\caption{Excess current density in the an-isotropic  reservoirs and the SBHZ wire. The anisotropy is introduced by choosing different hopping, denoted by $\eta_{bx}$ and $\eta_{by}$, in the $x$  and $y$ directions respectively. The current in the reservoirs is localized along the lines at $45$ degrees to the horizontal directions. Parameter values--$N_x=30$, $N_y=20$, $\Delta\mu_R=0$, $\eta_{bx}=\eta_c=1$, $\mu=0$ and $\mu_w=1$.}
		\label{fig:cden_an}
	\end{figure*}

The existence of such states at zero Fermi level and the current injection near the corners are the reasons for the observed localization of the current in the leads. The injection/ejection at the corner can be argued as follows: In the limit $W\rightarrow\infty$, by  current conservation and translation symmetry we expect no injection of the current across the junction. In the next section, Sec.~(\ref{sec:mci_jun}), we  construct scattering states of the metal-CI junction in the limit $N_y\rightarrow\infty$  to see this explicitly.  In the strip geometry, the translational symmetry is broken. However, we expect that the solutions of the scattering states to still hold away from the corners. This implies the injection of current  only occurs near the corners. For now, we present the numerical results on how the shape of the Fermi surface and disorder effect the observed current density patterns.

\begin{figure*}
	\subfigure[]{\includegraphics[width=0.881\textwidth]{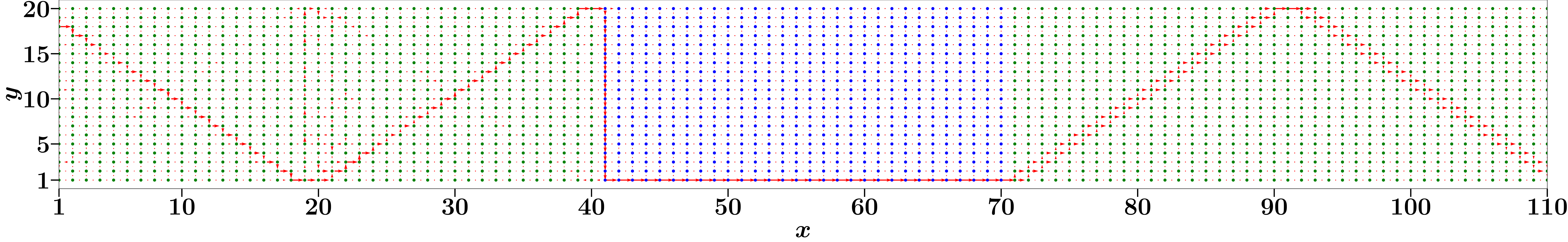}}
	\subfigure[]{\includegraphics[width=0.35\textwidth]{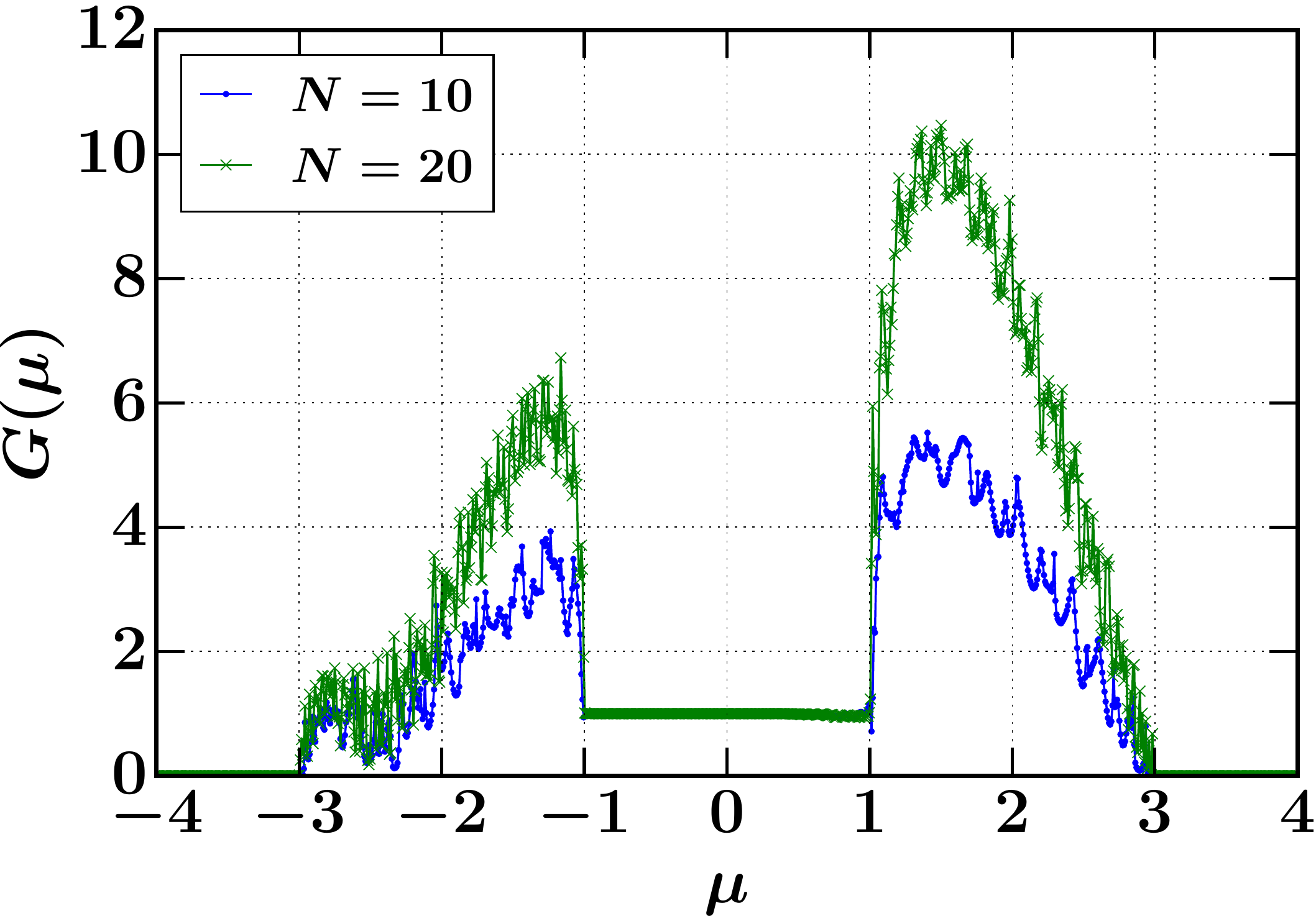}}
	\caption{(a) Excess current density in the   reservoirs and the SBHZ wire with disordered contacts. The disorder at the contacts is introduced by choosing the coupling, $\eta_c$, at every bond random from the interval $(0.75,1.25)$.  (b) shows the effect of disorder on the quantization of the two terminal conductance where the coupling, $\eta_c$ is chosen randomly in the interval $(1,2)$. Parameter values for (a) $N_x\times N_y=30\times20$, $\Delta\mu_R=0$, $\eta_{bx}=1$, $\mu=0$ and $\mu_w=1$ and for (b) $N_x=N_y=N$, $\eta_b=2$.}
		\label{fig:cden_do}
\end{figure*}

\subsection{Effects of the shape of the Fermi surface and disorder on current density}

If the Fermi surface is slightly deformed,  the features of the current density, namely the corner ejection and injection and localization inside the reservoirs, are retained up to slight perturbations. To illustrate this, we consider deforming the Fermi level in two ways: (a) tuning the Fermi level of the reservoirs away from zero and (b) introducing anisotropy inside the reservoirs while keeping the Fermi level at zero.  Fig.~(\ref{figsupp:cden1}) shows the current density at  $\mu=0.2$. The localization of the current  disappears deep in the metallic region, however,  the corner injection and ejection is retained.

We  introduce anisotropy  in the reservoirs by choosing different hoppings in the $x$ and $y$ directions,  respectively. The results are shown in Fig.~(\ref{fig:cden_an}), where we plot the current densities for $\eta_{by}=1.1$,  $\eta_{by}=1.2$ and $\eta_{by}=1.3$ with $\eta_{bx}=1$.  While the injection/ejection is still near the corners, the localization feature is retained for small perturbations but slowly gets washed away for increased anisotropy. Interestingly, on changing $\eta_{by}$, the angle of the lines with the $x$-axis also gets changed from $45$ degrees.

Topological quantities are expected to be robust to the effects of weak  disorder. Hence a natural question to investigate is as to what happens to the current density patterns and the quantized two-terminal conductance on introducing contact disorder.  We show the results in Fig.~\ref{fig:cden_do} where the coupling parameters between the system and the reservoir are chosen randomly. We see that  features of the current are almost unaffected by the disorder. This indicates that there may be a topological argument for these observed effects.

\section{Analytical Solution for Metal-Chern Insulator Junction}
\label{sec:mci_jun}
In this section, we  construct  explicit solutions for the  wave functions at zero energy in the limit $N_y\rightarrow\infty$ of a metal-CI junction to see explicitly that no current injection occurs into the CI. This is expected as translational symmetry in the limit $N_y\rightarrow\infty$  implies the current along the edges of the CI is constant. Hence,  a non-zero injection into the CI would violate current conservation. We argue that the solutions for zero energy wave functions constructed here approximate the solutions in the strip geometry away for the corners, thereby give no injection into the CI away from the corners. 

To simplify the problem, we ignore the right lead and consider only the left metallic lead in contact with the SBHZ Chern insulator with $N_y\rightarrow \infty$.   The Hamiltonian for the metallic sites is chosen to be the same as the left reservoir with all the hoppings set to one. The equation for the eigenstates of zero energy away from the junction between the metal and CI  are given by,  
	\begin{align}
		&\Psi_M(x+1,y)+\Psi_M(x-1,y)\notag\\&~~~~~+\Psi_M(x,y+1)+\Psi_M(x,y-1)=0;~x\leq -1,\label{eq:meqpos}\\
		&\frac{\sigma_z-i\sigma_x}{2}\Psi(x-1,y)+\frac{\sigma_z+i\sigma_x}{2}\Psi(x+1,y)+\mu_w\sigma_z \Psi(x,y)
		\notag\\&+\frac{\sigma_z-i\sigma_y}{2}\Psi(x,y-1)+\frac{\sigma_z+i\sigma_y}{2}\Psi(x,y+1)=0;~x\geq 2\label{eq:cieqpos},
	\end{align}
where $\Psi_M(x,y)$ and $\Psi(x,y)$ are the two component wave functions in the metal and inside the Chern insulator, respectively. Using the translation symmetry in the $y$-direction, we can substitute
\begin{align}
	\Psi_M(x,y)&=e^{ik y}\Phi_M(x),~\text{and}\\
	\Psi(x,y)&=e^{ik y}\Phi(x),\label{eq:cif_wfun}
\end{align}
into the Eq.~(\ref{eq:meqpos}) and Eq.~(\ref{eq:cieqpos}), respectively. This gives the equation for $\Phi_M(x)$ and $\Phi(x)$ to be,
\begin{align}
	&\Phi_M(x+1)+\Phi_M(x-1)+2\cos k \Phi_M(x)=0,\label{eq:mkspace}\\
	&\frac{\sigma_z-i\sigma_x}{2}\Phi(x-1)+\frac{\sigma_z+i\sigma_x}{2}\Phi(x+1)\notag\\&+\left[\mu_w\sigma_z 
	+\frac{\sigma_z-i\sigma_y}{2}e^{-ik}+\frac{\sigma_z+i\sigma_y}{2}e^{ik}\right]\Phi(x)=0.\label{eq:cikspace}
\end{align}
The zero energy solution inside the Chern insulator is localized so we substitute,
\begin{equation}
	\Phi(x)=\begin{pmatrix}
		u\\ v
	\end{pmatrix}z^{x-1},\label{eq:ansatz}
\end{equation}
in Eq.~\ref{eq:cikspace} and look for solution with $|z|<1$. This gives the  equation for $u$, $v$  to be
$
M\begin{pmatrix}
	u\\ v
\end{pmatrix}=0,
$ where 
\begin{equation}
	M=\begin{pNiceMatrix}
		\frac{1}{2}(z+\frac{1}{z})+\mu_w+\cos k & \frac{i}{2}(z-\frac{1}{z})+i \sin k\\
		\frac{i}{2}(z-\frac{1}{z})-i \sin k &-\left(\frac{1}{2}(z+\frac{1}{z})+\mu_w+\cos k\right)
	\end{pNiceMatrix}.
\end{equation}
For a non-trivial solution we have $\det[M]=0$ which  gives two solutions for $z$,
\begin{equation}
	z_\pm(k)=-\alpha(k)\pm \sqrt{\alpha^2(k)-1};~~\alpha(k)=\frac{\mu_w^2+2 \mu_w \cos k+2}{2(\mu_w+\cos k)}.
\end{equation}
Clearly, $z_+z_-=1$, so we choose the one with $|z|<1$ which for now we assume to be $z_+$. Using this we have for $u= \beta(k) v$ where,
\begin{align}
	\beta(k)=-\frac{-\alpha(k)+\mu_w+\cos k}{i (\sqrt{\alpha^2(k)-1}+ \sin k)}.
\end{align}

We can now write the full solution for the wave function at zero energy for the system as follows:
\begin{align}
	\Phi_M(x)&= e^{i k_x (x-1)} \begin{pmatrix} 1\\ B\end{pmatrix} +e^{-ik_x (x-1)} \begin{pmatrix}r_1\\r_2
	\end{pmatrix},\label{eq:wfm} \\
	\Phi(x)&=\begin{pmatrix}
		u\\\beta(k) u 
	\end{pmatrix}z^{x-1},\label{eq:wfci}
\end{align}
where $B$ is the amplitude of the incoming electron wave, and $r_1$, $r_2$ are its reflection coefficients onto the two degrees of freedom in the metal, respectively. 
To satisfy Eq.~(\ref{eq:mkspace}) inside the metal we take $k_x=k+\pi$. The coefficients $B$, $r_1$, $r_2$, and $u$  are chosen so that they satisfy the equations near the junction at $x=0$ and $x=1$. From the equations at these two values of $x$, we would basically get four linear equations for $u, B, r_1$ and $r_2$. We have at $x=0$,
\begin{align}
	\Phi_M(-1)+2\cos k \Phi_M(0)+\eta_c\Phi(1)=0\label{eqn1},\\
	\implies -\Phi_M(1)+\eta_c \Phi(1)=0\label{eqn2}.
\end{align}
While going from Eq.~\ref{eqn1} to Eq.~\ref{eqn2}, we used the fact that sum of the first, the second term in Eq.~(\ref{eqn1}) and $\Phi_M(x)$ evaluated at $x=1$ vanishes. We now have two linear equations for the coefficients $r_1,r_2, B$ and $u$. To obtain the other two equations, we consider the equation for the site at $x=1$. This is given by

\begin{align}
	&\eta_c\Phi_M(0)+\frac{\sigma_z+i\sigma_x}{2}\Phi(2)\notag\\&+\left[\mu_w\sigma_z +
	\frac{\sigma_z-i\sigma_y}{2}e^{-ik}+\frac{\sigma_z+i\sigma_y}{2}e^{ik}\right]\Phi(1)=0,\label{cieq1}\\
	\implies&\eta_c \Phi_M(0)-
	\left(\frac{\sigma_z-i\sigma_x}{2}\right)\Phi(0)=0.\label{cieq2}
\end{align}
	Eq.~(\ref{cieq2}) follows from the Eq.~(\ref{cieq1}) by noting that the sum of the second term, the third term in Eq.~(\ref{cieq1}) and the term $(1/2)(\sigma_z-i\sigma_x)\Phi(0)$ vanishes.  We therefore have the following linear equations for the coefficients $B$, $r_1$, $r_2$, and $u$ from Eq.~(\ref{eqn2}) and Eq.~(\ref{cieq2}):
	\begin{equation}
		\begin{pmatrix}
			0 && -1 && 0 && \eta_c\\
			-1 &&0 && -1 &&\beta\eta_c\\
			0 && -\eta_c e^{ik}&&0 && -g\\
			-\eta_c e^{-ik}&&0 &&-\eta_ce^{ik}&& i g
		\end{pmatrix}\begin{pmatrix}
			B\\r_1\\r_2\\ u
		\end{pmatrix}= \begin{pmatrix}
			1\\0\\\eta_c e^{-ik}\\0 
		\end{pmatrix},
	\end{equation} 
	where $g=(1-i \beta)/(2 z_+)$. From this linear system, the solution for the coefficients is given by,
	\begin{equation}
		\begin{pmatrix}
			B\\r_1\\r_2\\ u
		\end{pmatrix}	=\begin{pmatrix}
			\beta -\frac{g(i+\beta)}{g+\eta_c^2  e^{i k}}\\ -\frac{g+\eta_c^2  e^{-i k}}{g+\eta_c^2  e^{i k}}\\ \frac{i g-\beta  \eta_c^2  e^{-i k}}{g+\eta_c^2  e^{i k}}\\\frac{2 i \eta_c\sin k }{g+e^{ik}\eta_c^2}
		\end{pmatrix}.\label{eq:coffs}
	\end{equation}
	The current flowing in the $x$-direction from $x=0$ and $x=1$  between the first degree of freedom on the two sites is given by,
	\begin{equation}
		J_x^{11}=-2 I \eta_c \left[ \Phi^{*1}(1)\Phi_M^1(0)-\Phi^1(1)\Phi_M^{*1}(0)\right],
	\end{equation}
	where the superscript $1$ on $\Phi(0)$ and $\Phi_M(-1)$ refers to the first component of these two vectors. On simplification the current gives,
	\begin{align}
		J_x^{11}=-2i\eta_c(u^* (-e^{-i k }-e^{ik }r_1)-\text{c.c.}),\\
		= \frac{-8i\eta_c^2 }{|g+e^{ik}\eta_c|^2}(g-g^*)=0,
	\end{align}
	as $g$ is real. A similar calculation gives the current flowing between the other two degrees of freedom $J_x^{22}$ to be zero, as expected. 
	
	We now look at the  current in the y-direction at $x=1$ on the edge of the CI. This is given by
	\begin{equation}
		J_y= -2i (\Psi^\dagger(x,y) \frac{1}{2}(\sigma_x- i\sigma_y)  \Psi(x,y+1)-\text{c.c.})
	\end{equation} 
	where $\Psi(x,y)$ is given by Eq.~(\ref{eq:cif_wfun}). On simplification this gives a non-zero current,
	\begin{equation}
		J_y=8 \eta_c^2\frac{(1+\beta^2)\sin k-2\beta\cos k}{|g+\eta_c^2 e^{ik}|^2}\sin^2 k,
	\end{equation}
	which flows along the edge of the CI.
	
	While the solution given in Eq.~(\ref{eq:wfm}) and Eq.~(\ref{eq:wfci}) with the coefficients given by Eq.~(\ref{eq:coffs}) is a valid solution, this is not the only possible solution at a particular value of $k$. There exists another linearly independent solution for the same $k$. To see this, we consider a solution in the Chern insulator such that $(\sigma_z-i \sigma_x)\Phi(1)=0$. This condition makes $\Phi(1)$ vanish from the equations inside the Chern insulator. Therefore, the solution of the form chosen in Eq.~(\ref{eq:ansatz}) is no longer valid for $x=1$. Hence, at $x=1$ we are free to choose $\Phi(1)=\begin{pmatrix}
		1\\-i
	\end{pmatrix}\tilde{u}$, and for $x\geq 2$ the solution is still described by the ansatz in Eq.~(\ref{eq:ansatz}). At the boundaries, we now have four equations given by Eq.~(\ref{eqn2}) and Eq.~(\ref{cieq1}). For this case, Eq.~(\ref{cieq1}) should be used at the boundaries not  Eq.~(\ref{cieq2}). The latter is no longer true as $\Phi(1)$ is no longer given by the ansatz in Eq.~(\ref{eq:ansatz}). We therefore have four equations at the boundaries but five variables given by $B$, $r_1$, $r_2$, $\tilde{u}$ and $u$. The extra degree of freedom implies two linearly independent solutions for each $k$. If we choose $\tilde{u}=u$, then we get the solutions derived earlier with $\beta(k)=-i$. Another choice is to set $u=0$, then the equations at the junction give the following solution for the remaining coefficients, 
	\begin{equation}
		\begin{pmatrix}
			B\\r_1\\r_2\\u
		\end{pmatrix}=\begin{pmatrix} -i \left(1-\frac{g_-+g_+}{g_+-\eta_c ^2 e^{i k}}\right)\\-\frac{g_+-\eta_c ^2 e^{-i k}}{g_+-\eta_c ^2 e^{i k}}\\-\frac{i \left(g_-+\eta_c ^2 e^{-i k}\right)}{g_+-\eta_c ^2 e^{i k}}\\\frac{-2\eta_c  \sin k}{g_+-\eta_c^2e^{ik}}\end{pmatrix},
	\end{equation}
	where $g_\pm=\mu_w+\cos k \pm \sin k$. A similar calculation as earlier gives that at the junction the current in the $x-$direction vanishes while simultaneously the current in the $y-$direction is nonzero.  In the strip geometry, the translational symmetry is broken and therefore  allows the possibility of injection into the CI. However, we expect the solutions in the strip geometry to be similar to the above solutions away from the corners, and therefore do not give injection into the CI away from the corners. Thus current injection takes place only near the corners.

\section{Conclusions}
\label{sec:concl}
 In conclusion, we examined the assumptions of the LB formalism from a microscopic approach. These assumptions are what underlie the proof for the quantization of the two terminal conductance in 2D topological systems, but their validity is taken for granted. To understand why these assumptions are true, we looked at electronic transport properties due to the edge modes of a   Chern insulator (CI) in the open system geometry  using a microscopic approach based on the NEGF formalism,  applied to the spinless-BHZ model.  We found very nontrivial effects on the current density pattern inside the leads arising from the  topology of the insulator.  Particularly, our numerical results indicate that the current is highly localized, both in the metal and in the insulator, and the injection (ejection) into (from) the insulator happens only near the corners. The corner injection is robust to contact disorder and changes in the chemical potential. This provides a justification for the main assumption of LB, that of the emergence of an ideal  point contact  and consequently for the quantization of $G$. By the analysis of  scattering states that are formed at metal-CI interfaces we  have provided  some analytic arguments that support our observations of the current patterns.  
 
  While here we get the injection/ejection at diagonally opposite corners  in a quantum system which shows Hall effect type of physics, it is interesting to note that a completely classical treatment of Hall systems in the strip geometry, also gives rise to injection/ejection at diagonally opposite corners~\cite{PhysRevB.23.6610,prangequantum,rikken1988two,cage1997current} and this has been observed experimentally~\cite{dominguez1989optimization}. However, for obvious reasons, the classical calculations do not show the localization of current in either the CI or the metals.

We also verified  numerically  that the two terminal longitudinal conductance  in the open geometry is quantized and found very interesting finite-size effects.  
The conductance, at Fermi level $\mu=0$, grows non-monotonically with increasing metal-CI coupling strength, $\eta_c$, and increasing transverse system size, $N_y$. The growth to the quantized value shows oscillations with a period that scales as ${N_y}^{1/2}\eta_c$.

Analytic proofs of the quantization of the two-terminal conductance and the emergence of the ideal point contact for a large class of  models of insulators with non-trivial topology is desirable and remains an open problem. This would lead to an understanding of topology of open systems. Localized current patterns in the leads have  been observed in  other mesoscopic systems such as quantum point contacts~\cite{topinka2000imaging,topinka2003imaging,berezovsky2010imaging,chang2017nanoscale,tetienne2017quantum} and hence it should be possible to verify our predictions of localized currents in the leads. 
Our predictions of finite size-effects, in particular the oscillatory dependence on sample width and coupling, are also experimentally testable. 

\section{Acknowledgments}
We thank Adhip Agarwala, Anindya Das, Sriram Ganeshan, Joel Moore, Arun Paramekanti, Diptiman Sen, Shivaji Sondhi and Saquib Shamim for helpful discussions. 
J.M.B and A.D. acknowledge the support of the Department of Atomic Energy, Government of India, under Project No. RTI4001.

\appendix

\section{Proof, using the Landauer-Buttiker formalism, of equivalence between $G$ and $G_{H}$ in a 4-terminal Hall Bar setup}
\label{app:equi}

Here we show that use of the Landauer-Buttiker formalism, with the assumptions of single one-dimensional conducting channels and perfect contact resistances, gives us a proof of the equivalence of the two-terminal conductance, $G$, and the hall conductance, $G_H$.  
\begin{figure}[h!]
	\includegraphics[width=0.4\textwidth]{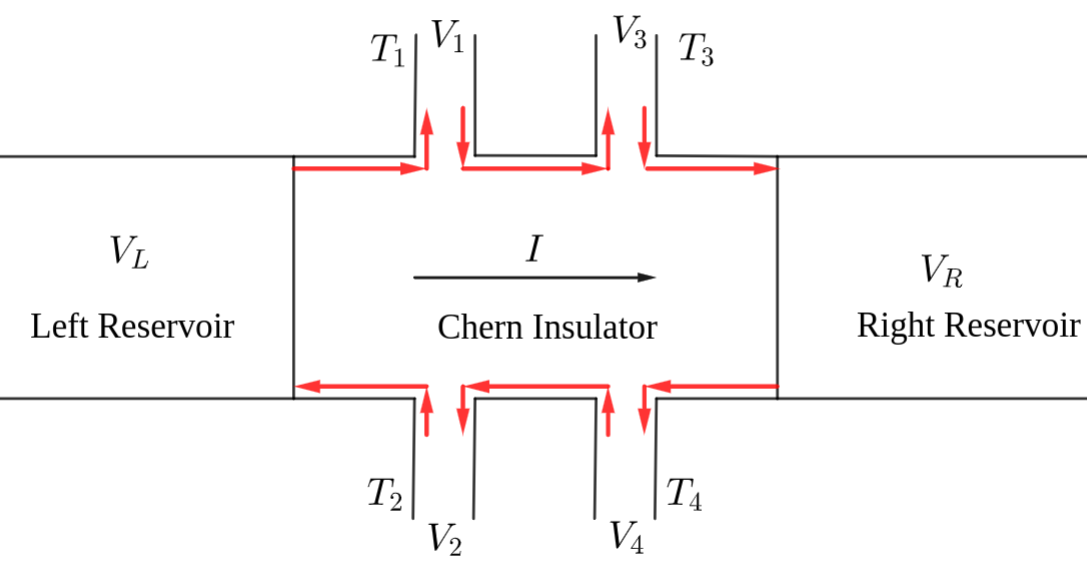}
	\caption{4-Terminal Hall bar setup.}
	\label{hallbar}
\end{figure}
Consider the 4-terminal hall bar setup shown in Fig.~(\ref{hallbar}). A current $I$ flows through the insulator from the left to the right reservoir. The definitions of two-terminal conductance, $G$ and the Hall conductance, $G_H$ are given by $G=I/(V_L-V_R)$ and  $G_H=I/(V_1-V_2)$. Under Landauer-Buttiker formalism and assuming that the chirality of the CI to be clockwise (indicated by red arrows in Fig.~(\ref{hallbar})), the currents flowing into the terminals $T_1$, $T_3$, $T_4$ and $T_2$ are given by,
\begin{align}
&I_1=\frac{e^2}{h}(V_1-V_L),~I_3=\frac{e^2}{h}(V_3-V_1),\notag\\&I_4=\frac{e^2}{h}(V_4-V_R)~\text{and}~I_2=\frac{e^2}{h}(V_2-V_4).
\end{align}

If no current  is allowed to flow into the probes, we have $V_L=V_1=V_3$, ans $V_2=V_4=V_R$, which immediately imply $G=I/(V_L-V_R)=I/(V_1-V_2)=G_H$.  The quantization of $G_H$ and $G$ is straightforward as under the assumption of perfect transmission of the Landauer-Buttiker formalism we have $I=\frac{e^2}{h}(V_L-V_2)=\frac{e^2}{h}(V_L-V_R)$. This directly gives $G_H$ as well as $G$ to be quantized to $e^2/h$. Note that the longitudinal resistance defined as $R_L=(V_1-V_3)/I$ vanishes, while $R_H=1/G_H$ is non-zero, and  this implies that the longitudinal conductance $G_L$ also vanishes (on inverting the resistance matrix).

\bibliography{biblio}
\end{document}